# Nipah virus vector sequences in COVID-19 patient samples sequenced by the Wuhan Institute of Virology

Steven C. Quay[1*], Daoyu Zhang[2], Adrian Jones[3] and Yuri Deigin[4].

[1] Atossa Therapeutics, Inc., Seattle, WA USA; ORCID 0000-0002-0363-7651

[2] Independent Genetics Researcher, Sydney, Australia

[3] Independent Bioinformatics Researcher, Melbourne, Australia

[4] Youthereum Genetics Inc., Toronto, Ontario, Canada; ORCID 0000-0002-3397-5811

[*] Correspondence to: Steven@DrQuay.com

## Abstract

We report the detection of Nipah virus in an infectious clone format, a BSL4-level pathogen and CDC-designated Bioterrorism Agent, in raw RNA-Seq sequencing reads deposited by the Wuhan Institute of Virology (WIV) produced from five December 2019 patients infected with SARS-CoV-2. Research involving Nipah infectious clones has never been reported to have occured at the WIV. These patient samples have been previously reported to contain reads from several other viruses: Influenza A, Spodoptera frugiperda rhabdovirus and Nipah. Previous authors have interpreted the presence of these virus sequences as indicative of co-infections of the patients in question by these pathogens or laboratory contamination. However, our analysis shows that NiV genes are encapsulated in synthetic vectors, which we infer was for assembly of a NiV infectious clone. In particular, we document the finding of internal N, P/V/W/C and L protein coding sequences as well as coverage of the G and F genes. Furthermore, the format of Hepatitis D virus ribozyme and T7 terminator downstream of the 5' end of the NiV sequence is consistent with truncation required at the end of the genome for a full length infectious clone. This indicates that research at WIV was being conducted on an assembled NiV infectious clone. Contamination of patient sequencing reads by an infectious NiV clone of the highly pathogenic Bangladesh strain could indicate a significant breach of BSL4 protocols. We call on WIV to explain the purpose of this research on infectious clones of Nipah Virus, the full chronology of this work, and to explain how and at what stage of sample preparation this contamination occurred.

# Introduction

Here we document the presence of Nipah virus (NiV) sequences, Banglisash strain, interpreted as likely for assembly of a NiV infectious clone, found in raw sequencing reads by the Wuhan Institute of Virology (WIV) from five patients infected with SARS-CoV-2 sampled by the Wuhan Jin Yin-Tan Hospital at the beginning of the COVID-19 outbreak (Zhou et al. 2020). The five patients experienced COVID-19 illness onset between 12/12/2019 and 23/12/2019 and were admitted to intensive care between 20/12/2019 and 29/12/2019 with all BALF (bronchoalveolar lavage fluid) sampling conducted on the 30/12/2019 and the 10/1/2020.

BioProject PRJNA605983 containing the analysed samples was registered with GenBank on 11/02/2020 and consists of nine RNA sequencing (RNA-Seq) BALF datasets. NGS (next-generation sequencing) was undertaken at the WIV using BGI MGISEQ-2000 and Illumina MiSeq 3000 sequencers (Zhou et al. 2020). Samples from patient tests WIV02, WIV04, WIV06 and WIV07 were sequenced on both platforms. While a sample from test name WIV05 was only sequenced on a MGISEQ-2000RS machine. Two lanes on a single flowcell were used for MGISEQ-2000 platform sequencing, with all samples sequenced on Lane 4 of this flowcell exhibiting high levels of contamination (Quay et al. 2021). As patient test WIV05 was not sampled in the second sampling on the 10/1/2020 and was only sequenced using the MGISEQ-2000RS platform, we infer that the first sampling (30/12/2019) may have been sequenced on the MGISEQ-2000RS platform and the second sampling, which contained only trace levels of contamination (Supp. Data 1 sheet 1.1), sequenced on the Miseq platform.

Here we re-examine the raw data for the earliest COVID-19 patient samples, (datasets SRR11092059, SRR11092060, SRR11092061, SRR11092062 and SRR11092063 found in BioProject PRJNA605983), which have been previously reported to contain reads from several other viruses: Influenza A virus, *Spodoptera frugiperda* rhabdovirus and Nipah virus (Chakraborty, 2020b; Abouelkhair, 2020; Zhang 2020; Quay et al. 2021). Chakraborty (2020a) and Abouelkhair (2020) interpreted the presence of these virus sequences as indicative of co-infection of early Wuhan COVID-19 infected patients with these microbes. However Quay et al. (2021) discuss the presence of a sequence H7N9 Hemagglutinin A segment 4 gene found in a synthetic vector in these COVID-19 patient samples and found contamination the likely cause, while Zhang (2020) identified the presence of a NiV infectious clone in the datasets. Here we review the recovery of a partial infectious clone of a Bangladesh strain of Nipah virus, a BSL4 level virus, with sequences joined directly to synthetic vector DNA.

NiV is a bat-borne zoonotic paramyxovirus first identified in 1999 as the cause of an outbreak of what was initially thought to be encephalitis in Malaysia in 1998 (Chua et al. 2000). Fruit bats of the genus *Pteropus* have been identified as the main reservoir host (Yob et al. 2001; Luby 2006). Repeated spillovers to humans have been documented in Bangladesh (Nikolay et al. 2019; Epstein et al. 2020). Isolation of NiVs from *P. medius* bats from Bangladesh under BSL 4 conditions was reported by Anderson et al. (2019).

The NiV genome consists of a negative-sense, single-stranded RNA of c. 18.2 kb, encoding six structural proteins, nucleoprotein (N), phosphoprotein (P), matrix protein (M), fusion protein (F), attachment glyco-protein (G), and RNA polymerase protein (L) and three nonstructural proteins (V, W and C proteins) (Sun et al. 2018). The glycoprotein G interacts to mediate attachment to receptors ephrin-B2 (Bonaparte et al. 2005; Negrete et al. 2005) or ephrin-B3 (Negrete et al. 2006).

## Results

Using a similar methodology as Abouelkhair (2020) and Quay et al. (2021) we used fastv analysis to identify virus sequences in each RNA-seq dataset from five patient samples in PRJNA605983 (Zhou et al. 2020) (Supp. Data 1). Samples sequenced on Lane 4 of the MGISEQ-2000RS machine (WIV07-2 (SRR11092059), WIV06-2 (SRR11092060), WIV05 (SRR11092061), WIV04-2 (SRR11092062) and WIV02-2 (SRR11092063) exhibited high levels of contamination not found in samples sequenced on Lane 2 or samples sequenced on the Illumina MiSeq platform (Supp. Data 1).

We aligned each dataset for samples WIV07-2 (SRR11092059), WIV06-2 (SRR11092060), WIV05 (SRR11092061), WIV04-2 (SRR11092062, Lane 4 only) and WIV02-2 (SRR11092063, Lane 4 only) to selected virus genomes previously identified in the datasets using fastv, including Nipah virus from Bangladesh, complete genome (AY988601.1) using bowtie2 (Fig. 1; Supp. Data 1 sheet 1.3).

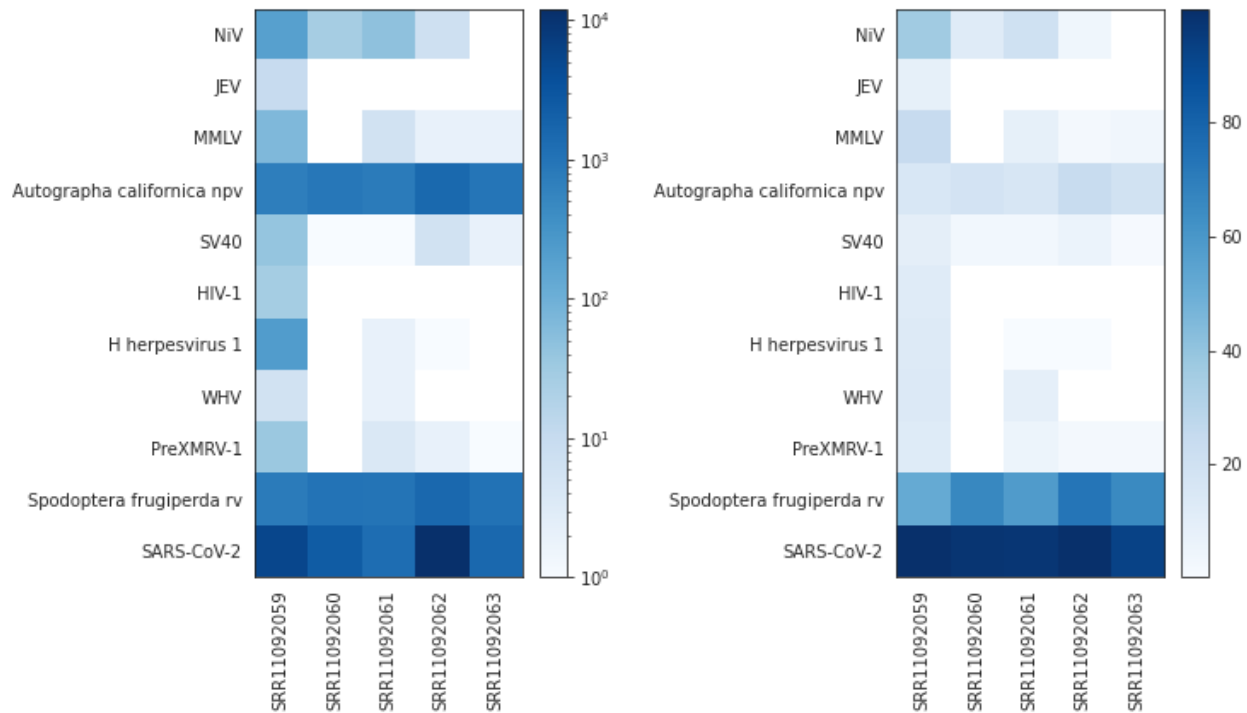

Fig. 1. Number of reads aligned (left, $\log_{10}$ scale) and percent genome coverage (right) for alignment of samples WIV07-2, WIV06-2, WIV05, WIV04-2 (Lane 4 only) and WIV02-2 (Lane 4 only) to selected virus genomes (see Supp. Data 1 sheet 1.4 for descriptions).

The RNA-Seq dataset for sample WIV07-2 exhibited highest read count for NiV and all other virus genomes analysed here with the exception of *Spodoptera frugiperda* rhabdovirus isolate Sf, *Autographa californica* nucleopolyhedrovirus and SARS-CoV-2. The higher contamination level in WIV07-2 than other samples was noted by Quay et al. (2021) with highest levels of pVAX1 plasmids containing Influenza A HA gene sequences in this patient sample.

*Nipah virus*

We pooled the datasets for WIV07-2, WIV06-2, WIV05, WIV04-2 (Lane 4 only) and WIV02-2 (Lane 4 only) for maximum NiV coverage and aligned to AY988601.1 using bowtie2 (Supp. Data 1 sheet 1.3) and bwa mem (Table 1). We generated a consensus sequence and analysed the sequence using NCBI BLASTN against the nt database using default parameters with highest max score (3984), total score (14071) and query coverage (42%) to Nipah virus from Bangladesh, complete genome (AY988601.1). All genes with the exception of the M gene (only 1nt coverage) had >100nt read coverage (Fig. 2). The non coding regions between N and P genes and P and M genes have complete or near complete coverage by reads, while the non coding region between the G and L genes has c. 50% read coverage (Fig. 2).

| Position | Identities | Percentage | Score bits | Expect value | Gaps | Gene(s) |
|---|---|---|---|---|---|---|
| 1-34 | 34/34 | 100% | 63.9 | 5E-04 | 0/34 | |
| 59-430 | 372/372 | 100% | 688 | 0.00 | 0/372 | N |
| 463-698 | 236/236 | 100% | 436 | 3E-116 | 0/236 | N |
| 763-2515 | 1752/1753 | 99% | 3234 | 0.00 | 0/1753 | N, P/V/W/C |
| 2571-4727 | 2157/2157 | 100% | 3984 | 0.00 | 0/2157 | P/V/W/C |
| 4767-5108 | 342/342 | 100% | 632 | 3E-175 | 0/342 | M (only 1nt) |
| 7964-8254 | 287/291 | 99% | 523 | 2E-142 | 0/291 | F |
| 8949-9138 | 190/190 | 100% | 351 | 1E-90 | 0/190 | G |
| 10758-11069 | 312/312 | 100% | 577 | 2E-158 | 0/312 | |
| 13137-13477 | 341/341 | 100% | 630 | 1E-174 | 0/341 | L |
| 14406 -14629 | 224/224 | 100% | 414 | 1E-109 | 0/224 | L |
| 16728-17622 | 868/895 | 97% | 1554 | 0.00 | 0/895 | L |
| 17699-18227 | 529/529 | 100% | 977 | 0.00 | 0/529 | L |

Table 1. BLASTN alignments of pooled 'MGISEQ L04' consensus sequence (derived from alignment to AY988601.1 using bwa mem) to Nipah virus from Bangladesh (AY988601.1).

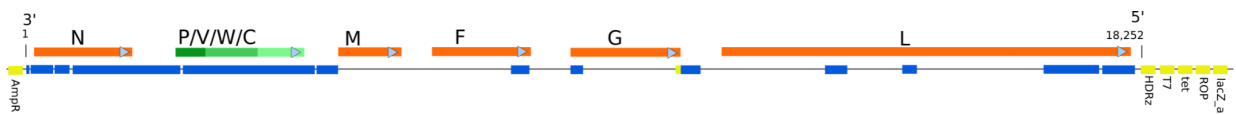

Fig. 2. BLASTN alignments of pooled 'MGISEQ L04' consensus sequence (derived from alignment to AY988601.1 using bwa mem) to Nipah virus from Bangladesh (AY988601.1). NiV genes in orange and green. Read coverage in blue, synthetic sequences in yellow (see also Supp. Data 2).

We ran a variant calling workflow on 'MGISEQ Lane 4' pooled reads using bcftools but did not detect any single nucleotide variants with a quality score >10%. Given that the coverage of the NiV by this pooled dataset is c. 42% and in particular coverage of the G and F gene sequences are low, it is possible that the NiV sequence is an unpublished, closely related strain to NiV Bangladesh (AY9688601.1).

We *do novo* assembled each of WIV07-2, WIV06-2, WIV05, WIV04-2 (Lane 4 only) and WIV02-2 (Lane 4 only) and a pool of these datasets ('MGISEQ Lane 4') using MEGAHIT and coronaSPAdes with similar results (Fig. 3).

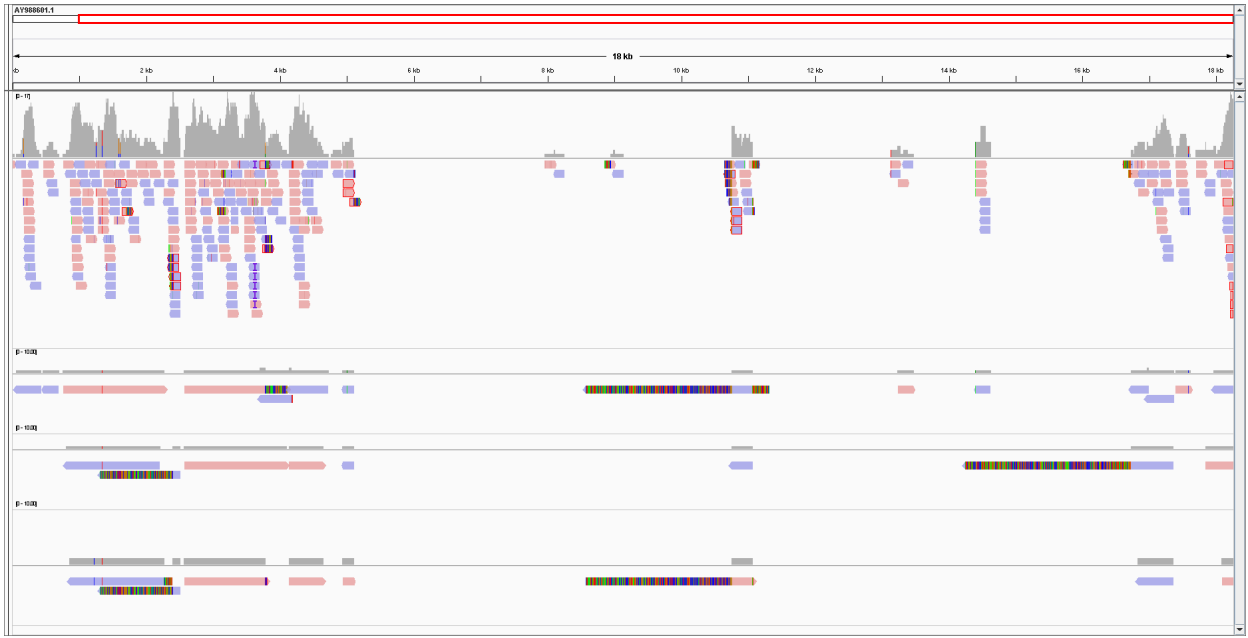

Fig. 3. Alignments to Nipah virus from Bangladesh (AY988601.1) for MGISEQ Lane 4 pooled reads (top track), coronaspades *de novo* assembled MGISEQ Lane 4 pooled reads (second track); MEGAHIT *de novo* assembled MGISEQ Lane 4 pooled reads (third track), MEGAHIT *de novo* assembled sample WIV07-02 (SRA11092059) using k79 mers (bottom track).

After *de novo* assembly of sample WIV07-2 with MEGAHIT we identified a contig covering the 5'-end of the NiV genome (1-166nt in vector sequence) fused to a *Hepatitis D virus* ribozyme, a T7 terminator, Tetracycline resistance gene, ROP gene, pGEX3 primer, lacZ reporter and M13 primers (Fig. 4, 5, Supp. Data 6).

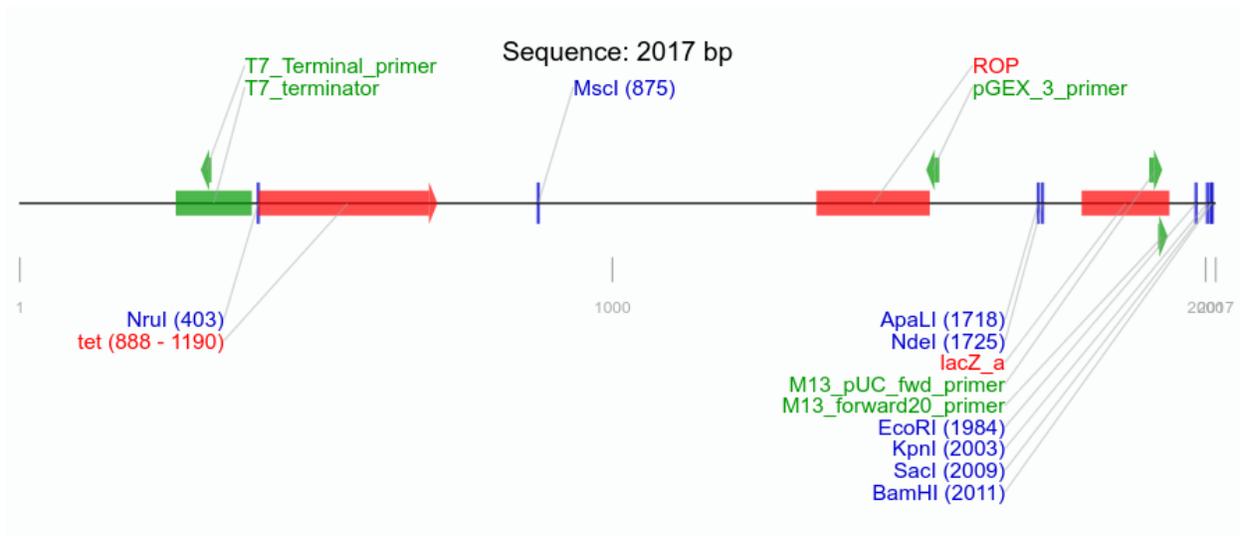

Fig. 4. Contig k141_119140 generated from *de novo* assembly of sample WIV07-2 (SRR11092059) containing NiV 5' end sequence at location 1-166nt.

Fig. 5. BLASTN alignment to the nt database for the first 166nt from contig k141_119140 from *de novo* assembled sample WIV07-2 (SRR11092059). A 100% match was found to L gene and trailer from 5' end of Nipah virus Bangladesh strain. See also Supp Fig. 1.

We additionally undertook *de novo* assembly using coronaSPAdes and identified a 2,743nt contig (2503nt when *de novo* assembled using k79 mer MEGAHIT assembly using SRR11092059 only (Supp. data 6)) that when analysed using BLASTN against the nt database exhibited a 1410 of 1411nt match to cloning vector pTAL7A. The synthetic sequence was attached to a 312nt section at the 3' end of the non coding region between the G and L genes (Fig. 3, Supp. Figs. 2, 3).

We identified further synthetic vector components from aligned read and contig analysis. Using the 'MGISEQ Lane 4' sequenced, pooled datasets aligned to Nipah virus from Bangladesh (AY988601.1), we used IGV to identify NiV aligned reads with >80nt contiguous soft-clipped ends. Eight synthetic sequences attached to NiV sequences were identified including cloning vectors and cloning or EGFP/luciferase reporter vector sequences (Supp. Data 3). A read from sample WIV07-2 was found to include an AmpR_promoter sequence upstream of the 3' end of the NiV sequence (Supp. Fig. 4), while a read covering part of the P/V/W coding region was found to contain a BGH reverse primer (Supp. Fig. 5). A partial reporter vector sequence was identified in a read at the 5' end of the non coding region between the P and M genes (Supp. Fig. 6). A read with a cloning vector attached to the 3' end of the G gene was found to have best matches to pcDNA3.1_+ like plasmids (Supp. Fig. 7). We note that reads with synthetic vector sequence ends in the N and P/V/W/C gene regions have synthetic sequences covered by other reads containing NiV sequences (Fig. 3).

*HIV-1*

Reads aligning to Human immunodeficiency virus 1, complete genome (NC_001802.1) were located in the rev/tat/env gene and nef gene regions. A consensus the 773nt and 159nt aligned regiones (Supp. Fig. 8) had a higher max score, total score and percentage identity to cloning and lentiviral vectors than to HIV-1 (NC_001802.1) (Supp. Fig. 9,10). Additionally, after *de novo* assembly and alignment of contigs to HIV-1 (NC_001802.1) a 308nt contig with a 271nt match to HIV-1 had a misaligned 37nt end with best match to numerous cloning vectors (Supp. Fig. 8, 11).

*SV40*

Reads in the 'MGISEQ Lane 4' pooled dataset were aligned to the Simian Virus genome (NC_001669.1) and found to align over only two small regions, a 328nt region covering the SV40 origin, promoter and enhancer region and a 152nt section mostly between the SV40gp5 and SV40gp6 genes (Supp. Fig. 12, 13). We generated a consensus for each region and analysed each against the nt database using BLASTN. The section covering the SV40 origin was best matched to multiple synthetic expression vectors (Supp. Fig. 14), while the 152nt section was found to best match numerous cloning vectors (Supp. Fig. 15).

*WHV*

Reads in the 'MGISEQ Lane 4' pooled dataset were aligned to the Woodchuck hepatitis virus, complete genome (NC_004107.1), with reads only covering part of the post-transcriptional regulatory element of the WHV genome. Sections of the open reading frame for the WHV X protein are covered (Supp. Fig. 16). A consensus of the aligned reads was analysed using BLASTN against the nt database with best matches to numerous synthetic vectors (Supp. Fig. 17).

## Discussion

NiV is classified as a BSL4 level pathogen due to its high case fatality rate of 32-41% for the Malaysia strain and c. 70% for the Bangladesh strain (Ang et al. 2018), and lack of an approved human vaccine. While Chakraborty (2020a) and Abouelkhair (2020) previously interpreted the discovery of NiV sequences in COVID-19 patient BALF samples as a NiV co-infection, here we document NiV genes contained in synthetic vectors, which we infer was for assembly of a NiV infectious clone. Significant contamination is evident throughout the MGISEQ-2000RS platform sequenced data, in particular sample WIV07-02 which contains c. 54% bacteria and higher Influenza A virus HA gene in pVAX1 plasmid sequence levels than SARS-CoV-2 reads (Quay et al. 2021; Supp Data 1 sheets 1.1, 1.2). Index hopping causing cross contamination during sequencing is largely homogeneous (Sinha et al. 2017, Ros-Freixedes et al. 2018) and the datasets studded here were all sequenced on the sample lane 4 on the same machine with flowcell id "@v300043428" (Supp. Data 1). However the level of NiV reads (and other contaminating sequences as discussed above) in sample WIV07-02 is significantly higher than other patient samples (Fig. 1; Supp Data 1 sheet 1.3), and we infer NiV contamination occurred prior to/or in the library preparation stage prior to sequencing (Cantalupo et al. 2019).

Pseudotyped NiV G and F envelope glycoproteins has been previously reported for HIV-1 (Palomares et al. 2013; Witting et al. 2013, Nie et al. 2019) and Vesicular Stomatitis Virus (Kaku et al. 2009; Kaku et al. 2012) backbone vector systems. While a minigenome containing the N, P and L genes were found to replicate efficiently (Halpin et. al. 2004, Freiberg et al. 2008) and like pseudotyped glycoproteins, could be studied under BSL-2 conditions. However, here we document the finding of internal N, P/V/W/C and L protein coding sequences as well as coverage of the G and F genes. Furthermore, the format of *Hepatitis D virus* ribozyme and T7 terminator downstream of the 5' end of the NiV sequence is consistent with truncation required at the end of the genome for a full length infectious clone. This indicates that research was being conducted on an assembled NiV infectious clone. However, there was virtually no read coverage for the M gene and we therefore cannot be certain if the full-length virus sequence was assembled.

Griffin et al. (2019) published the only known reverse genetics system for NiV-Bangladesh. Griffin et al. (2019) inserted marker gene (EGFP) between the G and L genes for *in vitro* NiV infection detection, used hammerhead ribozyme at the 3' end and a hepatitis delta virus ribozyme followed by a beta globin transcription terminator at the 5' end. Here we find a NannoLuc luciferase reporter sequence between the P and M genes. Griffin et al. (2019) discussed a preference for placing an EGFP ORF between the G and L genes rather than between the P and M genes to balance EGFP expression with preservation of viral polar transcription gradient and replication kinetics. Also differing from the system used by Griffin et al (2019), after the 5' end of the NiV sequence we find a hepatitis delta virus ribozyme followed by a T7 terminator rather than a beta globin transcription terminator. Further, a CMV promoter for RNA polymerase II and hammerhead ribozyme sequences Griffin et al. (2019) were not found at the 3' end of the NiV sequence found here.

NiV research was conducted as a priority at the WIV (Shi, 2019). However, after a search using Google Scholar and Pubmed, only two publications by WIV-affiliated authors were found in the 2018-2020 year period: a general overview of phylogeny, transmission and protein structure (Sun et al. 2018) and an article relating to rapid detection assay research, but which only concerns N gene pseudotyped NiV, rather than a NiV infectious clone (Ma et al. 2019).

We note the finding in relatively high abundance of reads matching *Spodoptera frugiperda* rhabdovirus isolate Sf, complete genome (NC_025382.1) in all RNA-seq datasets analysed here. *Spodoptera frugiperda* (Sf9) cells have been used previously for expressing the NiV M protein (Masoomi et al. 2016) and by WIV affiliated researchers for a baculovirus expression system for expressing NiV glycoprotein G (Yuan et al. 2011). The *Spodoptera frugiperda* rhabdovirus may be related to use of this cell line (Ma et al. 2014).

We also found *Autographa californica* nucleopolyhedrovirus (AcMNPV) in relatively high abundance in the studied datasets, which can infect Sf9 cells (Dong et al. 2010), and note that Sf9 cell entry and nucleocapsid assembly studies have been undertaken by the WIV (Wang et al. 2008; Qin et al. 2018).

Other contaminating sequences, including HIV-1, SV40 and WHV, were likely synthetic vector-related and not related to primary patient infection. Significant contamination at Wuhan sequencing facilities was previously documented by Zhang et al. (2021) with MERS-r CoV and SARS-r CoV genomes recovered from agricultural sequencing datasets. That sequences consistent with an infectious NiV clone, and numerous other synthetic sequences including a pVAX1 H7N9 HA plasmid (Quay et al. 2021) were found in samples from the earliest sequenced COVID-19 patients in Wuhan could indicate serious contamination problems at another sequencing site in Wuhan, the Wuhan Institute of Virology.

In summary, the finding of NiV gene sequences attached to synthetic vectors, presumably for assembly as a full length infectious NiV clone of the highly pathogenic Bangladesh strain, in RNA-Seq datasets for earliest sequences COVID-19 patients in Wuhan is potentially a significant breach of BSL4 protocols. We call on Zhou et al. (2020) to explain how and at what stage of sample preparation this contamination occurred. We further call on Zhou et al. to explain how abundant other contaminants including synthetic pVAX1 plasmids containing the HA gene of Influenza H7N9 (Quay et al. 2021) were found alongside NiV vector sequences is early COVID-19 patient samples sequenced by the WIV.

## Methods

All nine NGS datasets recording COVID-19 patient BALF sequencing in BioProject PRJNA605983 (Zhou et al. 2020): SRR11092056, SRR11092057, SRR11092058, SRR11092059, SRR11092060, SRR11092061, SRR11092062, SRR11092063, SRR11092064 were analyzed using the NCBI phylogenetic analysis tool STAT (Katz et al., 2021) (Supp. Data 1). Four datasets were identified as containing viruses unrelated to SARS-CoV-2 including Influenza A virus (IAV, subtype H7N9), *Spodoptera frugiperda* rhabdovirus and Nipah virus: SRR11092059, SRR11092060, SRR11092061, SRR11092062. A fifth contained Influenza A virus and *Spodoptera frugiperda* rhabdovirus: SRR11092063. All nine datasets were downloaded from NCBI. Fastv (Chen et al. 2020) was run for each SRA against the Opengene vial genome kmer collection.

The datasets were then subjected to BLAST search using MEGABLAST against the reference sequence of Nipah virus to verify the existence of the viral sequences and determine the exact subtype and the closest sequences on GenBank that corresponded to the reads. The Nipah virus genome was found to be AY988601.1.

MGISEQ sequenced datasets on lane 4 (SRR11092059,SRR11092060,SRR11092061, SRR11092062 (Lane04 only), SRR11092063 (Lane04 only)) were cleaned using fastp v0.20.1 (Chen et al. 2018). Each SRA was then aligned using bwa-mem v0.7.17 and or bowtie2 version 2.4.2 (Langmead and Salzberg, 2012) to selected virus genomes identified using fastv. Bamstats v1.25 and bamdst v1.0.9 were then used to calculate alignment statistics. GATK version 4.1.9.0 (Van der Auwera & O'Connor, 2020) was used to sort and mark duplicates. Samtools version 1.11 (Li et al. 2009) was used to index the marked .bam file for viewing in IGV version 2.8.13 (Thorvaldsdóttir et al. 2013).

We then undertook de novo assembly of each dataset in BioProject PRJNA605983 using MEGAHIT v1.2.9 (Li et al. 2015) and coronaSPAdes v3.15.2 (Meleshko et al. 2020). We further pooled all datasets sequenced on Lane04 of MGISEQ-2000RS platform id '@v300043428' and undertook *de novo* assembly of the combined dataset also using MEGAHIT and coronaSPAdes with default parameters.

We used bcftools v1.11 (Danecek et al. 2021) for variant calling on 'MGISEQ Lane 4' pooled reads.

A heatmap plot of virus reads and coverage was generated using matplotlib v3.3.4.

## Acknowledgements

We would like to thank Francisco de Asis for discussions of likely patient sequencing dates.

# Supplementary material

All supplementary material including the files listed below can be accessed at

DOI: 10.5281/zenodo.5515789

Link: https://zenodo.org/record/5515789

**Supp. Figs.**

Supp_Figs.pdf

**Supp. Data**

1_Datasets_and_Alignments.xlsx: Dataset details, fastv analysis results, alignment statistics, read counts and coverage.

2_AY988601_1_MGISEQ_L04_reads_MSA_view.png: Pooled reads from SRR11092059, SRR11092060, SRR11092061, SRR11092062 (Lane 4), SRR11092063 (Lane 4) aligned with bwa mem. Consensus generated in IGV then aligned using BLASTN to AY988601.1 and plotted in MSA viewer.

3_AY988601_1_MGISEQ_L04_reads_synthetic_ends_blast.fa: Identifiable synthetic ends to reads from pooled reads from SRR11092059, SRR11092060, SRR11092061, SRR11092062 (Lane 4), SRR11092063 (Lane 4) aligned with bwa mem.

4_AY988601_1_BLASTN_default_hits.fa: BLASTN of AY988601.1 against the following search set: SRR11092059, SRR11092060, SRR11092061, SRR11092062, SRR11092063 using max target sequences of 1000.

5_AY988601_1_MGISEQ_L04_MEGAHIT_contigs_all.sam: MEGAHIT de novo assembled contigs from SRR11092059, SRR11092060, SRR11092061, SRR11092062 (Lane 4), SRR11092063 (Lane 4), aligned with bwa mem to AY988601.1.

6_AY988601_1_SRR11092059_MEGAHIT_k79_contigs.sam: MEGAHIT de novo assembled k79 mer contigs from SRR11092059 aligned with bwa mem to AY988601.1.

7_AY988601_1_MGISEQ_L04_coronaSPAdes_contigs_all.sam: CornaSPAdes de novo assembled contigs from SRR11092059, SRR11092060, SRR11092061, SRR11092062 (Lane 4), SRR11092063 (Lane 4), aligned with bwa mem to AY988601.1.